\documentclass[letterpaper,10 pt,conference]{ieeeconf}
\IEEEoverridecommandlockouts                              
\usepackage[utf8]{inputenc}
\usepackage{amsmath}
\usepackage{mathrsfs}
\usepackage{amssymb}
\usepackage{color}
\usepackage{dsfont}
\usepackage{cite}
\usepackage{comment}
\usepackage{float}
\usepackage{graphicx}
\usepackage[font = footnotesize]{caption}
\usepackage[font = footnotesize]{subcaption}
\usepackage{goelStyle}
\usepackage{color}
\usepackage{xcolor}
\usepackage{float}
\usepackage{tikz}
\usepackage{booktabs}
\usepackage{tabularx}
\usepackage{multirow}
\usepackage{bm}
\usepackage{makecell}
\usepackage{indentfirst}
\newcolumntype{C}[1]{>{\centering\let\newline\\\arraybackslash\hspace{0pt}}m{#1}}

\usepackage{color, colortbl}
\definecolor{Yellow}{rgb}{1,1,0}

\usepackage{hhline}

\usetikzlibrary{shapes,arrows,calc,positioning,patterns}
\tikzstyle{bigblock} = [draw, fill=blue!20, rectangle, 
    minimum height=6em, minimum width=8em]
\tikzstyle{medblock} = [draw, fill=blue!20, rectangle, 
    minimum height=4em, minimum width=4em]    
\tikzstyle{mux} = [draw, fill=black!20, rectangle, 
    minimum height=5em, minimum width=0.1em]    
\tikzstyle{smallblock} = [draw, fill=blue!20, rectangle, 
    minimum height=3em, minimum width=4em]
\tikzstyle{sum} = [draw, fill=blue!20, circle, node distance=1cm]
\tikzstyle{signal} = [coordinate]
\tikzstyle{pinstyle} = [pin edge={to-,thin,black}]
\tikzstyle{block} = [draw, fill=blue!20, rectangle, 
    minimum height=3em, minimum width=6em]
\tikzstyle{blockS} = [draw, fill=blue!20, rectangle, 
    minimum height=3em, minimum width=4em]  
\tikzstyle{sum} = [draw, fill=blue!20, circle, node distance=1.5cm]
\tikzstyle{gain} = [draw, fill=blue!20, regular polygon, regular polygon sides = 3, node distance=1.25cm, shape border rotate = -90]
\tikzstyle{mult} = [draw, fill=blue!20, circle, node distance=1.25cm ,inner sep=0pt, minimum size = 0.3cm]

\tikzstyle{input} = [coordinate]
\tikzstyle{output} = [coordinate]
\tikzstyle{spring}=[thick,decorate,decoration={zigzag,pre length=0.3cm,post length=0.3cm,segment length=6}]
\tikzstyle{damper}=[thick,decoration={markings,  
  mark connection node=dmp,
  mark=at position 0.5 with 
  {
    \node (dmp) [thick,inner sep=0pt,transform shape,rotate=-90,minimum width=15pt,minimum height=3pt,draw=none] {};
    \draw [thick] ($(dmp.north east)+(2pt,0)$) -- (dmp.south east) -- (dmp.south west) -- ($(dmp.north west)+(2pt,0)$);
    \draw [thick] ($(dmp.north)+(0,-5pt)$) -- ($(dmp.north)+(0,5pt)$);
  }
}, decorate]
\tikzstyle{ground}=[fill,pattern=north east lines,draw=none,minimum width=0.75cm,minimum height=0.3cm]

\DeclareMathOperator*{\argmin}{argmin}

\newcounter{example}

\usepackage{hyperref}
\usepackage{xcolor}
\hypersetup{
    colorlinks,
    linkcolor={blue!100!black},
    citecolor={blue!50!black},
    urlcolor={blue!80!black}
}

\title{\LARGE Explicit Model Predictive Control \\ based on a Fuzzy-Autoregressive Moving Average Model}

\title{\LARGE Control Approximation based on Fuzzy ARMA Controller Applied to MPC}

\title{\LARGE Data-driven Fuzzy Logic-based Control Synthesis using MPC Framework}

\title{\LARGE MPC-guided, Data-driven Fuzzy Controller Synthesis}

\author{\large Juan Augusto Paredes Salazar and Ankit Goel
\thanks{Juan Augusto Paredes Salazar and Ankit Goel are with the Department of Mechanical Engineering, University of Maryland, Baltimore County, MD 21250.
{\tt \small \{japarede,ankgoel\}@umbc.edu}}
}

\begin{document}

\maketitle

\begin{abstract}
Model predictive control (MPC) is a powerful control technique for online optimization using system model-based predictions over a finite time horizon.
However, the computational cost MPC requires can be prohibitive in resource-constrained computer systems.
%
%
%
This paper presents a fuzzy controller synthesis framework guided by MPC.
%
In the proposed framework, training data is obtained from MPC closed-loop simulations and is used to optimize a low computational complexity controller to emulate the response of MPC.
%
%
In particular, autoregressive moving average (ARMA) controllers are trained using data obtained from MPC closed-loop simulations, such that each ARMA controller emulates the response of the MPC controller under particular desired conditions.
Using a Takagi-Sugeno (T-S) fuzzy system, the responses of all the trained ARMA controllers are then weighted depending on the measured system conditions, resulting in the Fuzzy-Autoregressive Moving Average (F-ARMA) controller.
The effectiveness of the trained F-ARMA controllers is illustrated via numerical examples.
\end{abstract}


\section{Introduction} \label{sec:introduction}

Model predictive control (MPC) is a feedback control framework that can simultaneously minimize a user-defined, system output-dependent cost function over a future finite time horizon and satisfy user-defined constraints \cite{grune2017,rakovic2018,schwenzer2021}.
%
%
However, the computational cost MPC requires for model prediction and subsequent optimization makes its implementation in resource-constrained applications prohibitive \cite{mesbah2016,saltik2018}.
%
%
A popular technique to address this issue consists of computing MPC offline and fitting a piecewise-affine function to the MPC input and output data, which makes the dependence of the MPC output on the MPC input explicit; hence, this technique is called Explicit MPC (EMPC) \cite{alessio2009}.
While this solution dramatically reduces the computational cost, the complexity of the piecewise-affine function grows as the state dimension and the nonlinearity of the system dynamics increase, which increase the computational and memory requirements for implementation \cite{wen2009,scibilia2009,jones2010,kvasnica2011,nguyen2017,kvasnica2019}.
An alternative to EMPC is given by MPC-guided control synthesis, also know as imitation learning, in which training data obtained from MPC closed-loop simulations is used to synthesize a low computational complexity controller that emulates the response of MPC, which usually take the form of neural networks \cite{zhang2016,hertneck2018,kaufmann2020,zhang2020,chen2021,tagliabue2022,tagliabue2024}.

In this work, an MPC-guided control synthesis procedure is used to synthesize a fuzzy controller that emulates the response of the original MPC controller.
The proposed synthesis procedure involves training autoregressive moving average (ARMA) controllers using data obtained from MPC closed-loop simulations, such that each ARMA controller emulates the response of the MPC controller under particular conditions.
For online implementation, the responses of all the trained ARMA controllers are weighted depending on the measured system conditions and interpolated using a Takagi-Sugeno (T-S) fuzzy system \cite[ch.~6]{lilly2011}, and the resulting controller is called Fuzzy-Autoregressive Moving Average (F-ARMA) controller.
The T-S fuzzy framework is chosen since it provides an intuitive methodology to interpolate the response of linear systems for control applications
\cite{nguyen2019,precup2024}.
%
In partiuclar, T-S fuzzy systems are used to interpolate the responses of linear dynamic models and controllers to increase the domain of attraction of control techniques, such as
LQR
\cite{leal2021,wang2022,al2023},
%
MPC
\cite{aslam2023,mendes2024,sayadian2024},
%
and neural network-based controllers
\cite{cervantes2020,pham2023,khater2024}.

The contents of this paper are as follows.
Section \ref{sec:prob} briefly reviews the sampled-data feedback control problem for a continuous-time dynamic system.
Section \ref{sec:MPC} reviews the nonlinear MPC (NMPC) and specializes this technique for linear plant, yielding linear MPC.
Section \ref{sec:F-ARMA} introduces the F-ARMA controller synthesis framework.
Section \ref{sec:LS} describes a least-squares regression technique to synthesize the ARMA controllers that compose F-ARMA using data obtained from closed-loop MPC simulations.
Section \ref{sec:numerical_examples} presents examples that illustrate the performance of the proposed F-ARMA algorithm and its effectiveness at emulating the response of MPC.
Finally, the paper concludes with a summary in Section \ref{sec:conclusions}.

{\bf Notation:}
$\BBR \isdef (-\infty,$ $\infty),$ 
$\BBR^{\ge0} \subset \BBR \isdef [0,$ $\infty),$ 
$\BBZ \isdef \{\ldots, -2, -1, 0, 1, 2, \ldots\},$
and $\Vert\cdot\Vert$ denotes the Euclidean norm on $\BBR^n.$
$x_{(i)}$ denotes the $i$th component of $x\in\BBR^n.$
The symmetric matrix $P\in\BBR^{n\times n}$ is positive semidefinite (resp., positive definite) if all of its eigenvalues are nonnegative (resp., positive).
$\vek X\in\BBR^{nm}$ denotes the vector formed by stacking the columns of $X\in\BBR^{n\times m}$, and 
$\otimes$ denotes the Kronecker product.
$I_n$ is the $n \times n$ identity matrix, $0_{n\times m}$ is the $n\times m$ zeros matrix, and $\mathds{1}_{n\times m}$ is the $n\times m$ ones matrix.
${\rm wrap}_\pi \colon \BBR \to (-\pi, \ \pi]$ wraps the input angle, such that ${\rm wrap}_\pi (x) \in (-\pi, \ \pi]$ for all $x \in \BBR.$



\section{Reference Tracking Control Problem}\label{sec:prob}

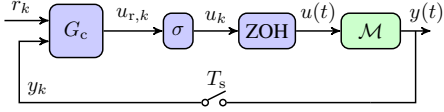
\begin{figure} [h!]
    \centering
    \vspace{-0.25em}
    \resizebox{0.7\columnwidth}{!}{%
    \begin{tikzpicture}[>={stealth'}, line width = 0.25mm]

    \node [input, name=ref]{};
    \node [smallblock, rounded corners, right = 0.5cm of ref , minimum height = 0.9cm, minimum width = 0.9cm] (controller) {$G_\rmc$};
    \node[smallblock, rounded corners, right = 1cm of controller, minimum height = 0.6cm, minimum width = 0.5cm](sat_Blk){$\sigma$};
    \node [smallblock, rounded corners, right = 0.75cm of sat_Blk, minimum height = 0.6cm , minimum width = 0.5cm] (DA) {ZOH};
    
    \node [smallblock, fill=green!20, rounded corners, right = 0.75cm of DA, minimum height = 0.6cm , minimum width = 1cm] (system) {$\SM$};
    \node [output, right = 0.5cm of system] (output) {};
    \node [input, below = 0.9cm of system] (midpoint) {};
    
    \draw [->] (controller) -- node [above] {$u_{\rmr,k}$} (sat_Blk);\
    \draw [->] (sat_Blk) -- node [above] {$u_k$} (DA);\
    \draw [->] (DA) -- node [above] {$u (t)$} (system);
    
    \node[circle,draw=black, fill=white, inner sep=0pt,minimum size=3pt] (rc11) at ([xshift=-2.5cm]midpoint) {};
    \node[circle,draw=black, fill=white, inner sep=0pt,minimum size=3pt] (rc21) at ([xshift=-2.8cm]midpoint) {};
    \draw [-] (rc21.north east) --node[below,yshift=.55cm]{$T_\rms$} ([xshift=.3cm,yshift=.15cm]rc21.north east) {};
    
    \draw [->] (system) -- node [name=y, near end]{} node [very near end, above] {$y (t)$}(output);
    
    \draw [-] (y.west) |- (midpoint);
    \draw [-] (midpoint) -| node [very near end, above, xshift=-3.2cm] {$y_k$} (rc11.east);
    \draw [->] (rc21) -| ([xshift = -0.5cm, yshift = -0.5em]controller.west) -- ([yshift = -0.5em]controller.west);
    \draw [->] ([xshift = -0.75cm, yshift = 0.5em]controller.west) -- node [above, xshift=-0.25em] {$r_k$} ([yshift = 0.5em]controller.west);
    
    \end{tikzpicture}
    }  
    \caption{Sampled-data implementation of discrete-time, reference tracking controller applied to a continuous-time system $\SM$ with input $u$ and output $y.$
    All sample-and-hold operations are synchronous.
    %
    }
    \label{fig:RT_CT_blk_diag}
    \vspace{-0.5em}
\end{figure}

To reflect the practical implementation of digital controllers for physical systems, we consider continuous-time dynamics under sampled-data control using a discrete-time, reference tracking controller.
In particular, we consider the control architecture shown in Figure \ref{fig:RT_CT_blk_diag}, where $\SM$ is the target continuous-time system, $t\ge 0$, $u(t)\in\BBR^{\ell_u}$ is the control, and $y(t)\in\BBR^{\ell_y}$ is the output of $\SM,$ which is sampled to produce the sampled output measurement $y_k \in \BBR^{\ell_y},$ which, for all $k\ge0,$ is given by
%
    $y_k \isdef  y(k T_\rms),$
%
where  $T_\rms>0$ is the sample time.

The discrete-time, reference tracking is denoted by $G_\rmc.$
The inputs to $G_{\rmc,k}$ are $y_k$ and the reference $r_k,$ and its output at each step $k$ is the requested discrete-time control $u_{\rmr,k}\in\BBR^{\ell_u}.$ 
Since the response of a real actuator is subjected to hardware constraints, the implemented discrete-time control is 
\begin{equation}
 u_k\isdef \sigma(u_{\rmr,k}),  
 \label{eq:ukksat}
\end{equation}
where $\sigma\colon\BBR^{\ell_u}\to\BBR^{\ell_u}$ is the control-magnitude saturation function  
\begin{equation}
    \sigma(u) \isdef  \matl \bar{\sigma}_1 (u_{(1)}) \\ \vdots \\ \bar{\sigma}_{\ell_u} (u_{(\ell_u)})\matr,
    \label{eq:Hsat1}
\end{equation}
where, for all $i \in \{1, \ldots, \ell_u \},$ $\bar\sigma_i \colon \BBR\to\BBR$ is defined by 
\begin{equation}
    \bar{\sigma}_i (u_{(i)}) \isdef
    \begin{cases} 
    u_{\max, (i)},&u_{(i)} > u_{\max, (i)},\\
    u_{(i)},& u_{\min, (i)} \le u_{(i)} \le u_{\max, (i)},\\
    u_{\min, (i)}, & u_{(i)} < u_{\min, (i)},
    \end{cases}
    \label{eq:Hsat2}
\end{equation}
and $u_{\min, (i)},u_{\max, (i)}\in\BBR$ are the lower and upper magnitude saturation levels corresponding to $u_{(i)},$ respectively.
Similarly, by appropriate choice of the function $\sigma, $
rate (move-size) saturation can also be considered.
The continuous-time control signal $u(t)$ applied to the system $\SM$ is generated by applying a zero-order-hold operation to $u_k,$ that is,
for all $k\ge0,$ and, for all $t\in[kT_\rms, (k+1) T_\rms),$ 
\begin{equation}
    u(t) = u_k.
\end{equation}

The objective of the discrete-time, reference tracking controller is to generate an input signal that minimizes 
$\sum_{k = 0}^{\infty} V(r_k, y_k),$ where $V \colon \BBR^{\ell_y} \times \BBR^{\ell_y} \colon \to \BBR^{\ge0}$ is a user-defined cost function whose complexity depends on the tracking objective and the nonlinear properties of $\SM.$
In most cases, the tracking objective is to minimize the norm of the difference between the sampled output measurement and reference signals, that is, $V(r_k, y_k) = \Vert r_k - y_k\Vert$ and thus $\sum_{k = 0}^{\infty} \Vert r_k - y_k\Vert$ is minimized.
This paper considers the problem of inverting a pendulum on a linear cart, 
where $r_k$ and $y_k$ correspond to reference and measured angles in radians, respectively.
The tracking objective of reaching an upwards position may be encoded in the cost function as $V(r_k, y_k) = 1 - \cos(y_k)$ in the case where the upwards position angle is given by $2 \pi i$ for all $i \in \BBZ$. 
%

In this paper, we use the MPC techniques reviewed in Section \ref{sec:MPC} to synthesize the controller $G_\rmc$ given by the F-ARMA controller, presented in Section \ref{sec:F-ARMA}.
In particular, we use MPC to generate a control sequence for trajectory tracking, and then use the measured data from the MPC-based controller to train the linear controllers in the F-ARMA framework to emulate the response of the original MPC controller.
%
%





\section{Overview of Nonlinear Model Predictive Control} \label{sec:MPC}

For all $k\ge 0,$ let the dynamics of a nonlinear, discrete-time system be given by
\begin{align}
    x_{k+1} &= f(x_k, u_k), \label{eq:NLD_x}\\
    y_k &= h(x_k, u_k),\label{eq:NLD_y}
\end{align}
where $x_k \in \BBR^{\ell_x}$ is the sampled state, $f \colon \BBR^{\ell_x} \times \BBR^{\ell_u} \to \BBR^{\ell_x},$ and  $h \colon \BBR^{\ell_x} \times \BBR^{\ell_u} \to \BBR^{\ell_y}.$
In this work, it is assumed that $f$ is at least twice continuously differentiable and that, for all $k \ge 0,$ $x_k$ and a state reference $x_{\rmr,k} \in \BBR^{\ell_x}$ can be obtained from $y_k$ and $r_k.$
Then, the NMPC problem to solve at each step $k > 0$ to minimize $\sum_{k = 0}^{\infty} V(r_k, y_k)$ is given by
\begin{align}
    &X_{{\rm mpc}, k} = \argmin_{X \in \BBR^{\ell_\rmh (\ell_x + \ell_u)}} V_{\rmf, x} (\overline{x}_{\rmr,\ell_\rmh}, \overline{x}_{\ell_\rmh}) \nn \\ 
    & \hspace{4em} + \sum_{i = 1}^{\ell_\rmh-1} V_x (\overline{x}_{\rmr,i}, \overline{x}_i) + \sum_{i = 0}^{\ell_\rmh-1}V_u (\overline{u}_i), \label{eq:NMPC_J}
\end{align}
subject to
\begin{align}
    &\overline{x}_1 = f(x_k, \overline{u}_0), \label{eq:NMPC_equal_0}\\
    &\overline{x}_{i + 1} = f(\overline{x}_i, \overline{u}_i) \mbox{ for all } i \in \{1, \ldots, \ell_\rmh-1 \}, \label{eq:NMPC_equal}\\
    &\matl I_{\ell_u} \\ -I_{\ell_u} \matr \overline{u}_{i} \leq \matl u_{\rm max} \\ u_{\rm min} \matr \mbox{ for all } i \in \{0, \ldots, \ell_\rmh-1 \}, \label{eq:NMPC_inequal}
\end{align}
where
\begin{equation*}
    u_{\min} \isdef  \matl u_{\min, (1)} \\ \vdots \\ u_{\min,(\ell_u)}\matr, \quad u_{\max} \isdef  \matl u_{\max,(1)} \\ \vdots \\ u_{\max,(\ell_u)}\matr \in \BBR^{\ell_u},
\end{equation*}
$\ell_\rmh > 0$ is the optimization horizon, 
$\overline{x}_1, \ldots, \overline{x}_{\ell_\rmh} \in \BBR^{\ell_x}$ are the state optimization variables,
$\overline{x}_{\rmr, 1}, \ldots, \overline{x}_{\rmr, \ell_\rmh} \in \BBR^{\ell_x}$ are the state reference optimization variables,
$\overline{u}_0, \ldots, \overline{u}_{\ell_\rmh-1} \in \BBR^{\ell_u}$ are the input optimization variables,
$V_{\rmf, x} \colon \BBR^{\ell_x} \times \BBR^{\ell_x} \to \BBR^{\ge0}$ is the terminal state cost function and $V_x \colon \BBR^{\ell_x} \times \BBR^{\ell_x} \to \BBR^{\ge0}$ is the state cost function, both of these are chosen to meet same tracking objective as $V,$
$V_u \colon \BBR^{\ell_u} \to \BBR^{\ge0}$ is the input cost function, chosen to discourage solutions with high-magnitude inputs,
$X \in \BBR^{\ell_\rmh (\ell_x + \ell_u)}$ encodes the optimization variables, such that
\begin{equation*}
    X \isdef \matl \overline{u}_0^\rmT & \cdots & \overline{u}_{\ell_\rmh -1}^\rmT & \overline{x}_1^\rmT & \cdots & \overline{x}_{\ell_\rmh}^\rmT \matr^\rmT,
\end{equation*}
and $X_{{\rm mpc}, k} \in \BBR^{\ell_\rmh (\ell_x + \ell_u)}$ encodes the NMPC solution at step $k,$ such that
\begin{equation*}
    X_{{\rm mpc}, k} \isdef \matl u_{\rms, 0}^\rmT & \cdots & \overline{u}_{\rms, \ell_\rmh -1}^\rmT & \overline{x}_{\rms, 1}^\rmT & \cdots & \overline{x}_{\rms, \ell_\rmh}^\rmT \matr^\rmT,
\end{equation*}
where $\overline{x}_{\rms,1}, \ldots, \overline{x}_{\rms, \ell_\rmh} \in \BBR^{\ell_x}$ are the state solution variables,
and $\overline{u}_{\rms, 0}, \ldots, \overline{u}_{\rms, \ell_\rmh-1} \in \BBR^{\ell_u}$ are the input solution variables.
Then, the control input at step $k$ is given by the initial control input solution, such that
\begin{equation}
    u_k = \overline{u}_{\rms, 0} = \matl I_{\ell_u} & 0_{\ell_u \times (\ell_\rmh \ell_x + (\ell_\rmh - 1) \ell_u)} \matr X_{{\rm mpc}, k}.\label{eq:uk_NMPC}
\end{equation}
In this work, we choose $\overline{x}_{\rmr,i} = x_{\rmr,k}$ for all $i \in \{1, \ldots, \ell_\rmh\}$ at each step $k>0$ to track constant reference, although other values could be chosen to accommodate complex trajectories.

The NMPC problem can be solved by rewriting \eqref{eq:NMPC_J}$-$\eqref{eq:NMPC_inequal} as a nonlinear program (NLP).
Then, the NLP to solve at each step $k > 0$ to minimize $\sum_{k = 0}^{\infty} V(r_k, y_k)$ is given by
\begin{equation}
    X_{{\rm mpc}, k} = \argmin_{X \in \BBR^{\ell_\rmh (\ell_x + \ell_u)}} J(X), \label{eq:NLP_J}
\end{equation}
subject to
\begin{align}
    & g_{\rm eq} (X) = 0, \label{eq:NLP_equal}\\
    & g_{\rm ineq} (X) \le 0, \label{eq:NLP_inequal}
\end{align}
where $J \colon \BBR^{\ell_\rmh (\ell_x + \ell_u)} \to \BBR^{\ge0},$ $g_{\rm eq} \colon \BBR^{\ell_\rmh (\ell_x + \ell_u)} \to \BBR^{\ell_\rmh \ell_x},$ and $g_{\rm ineq} \colon \BBR^{\ell_\rmh (\ell_x + \ell_u)} \to \BBR^{2 \ell_\rmh \ell_u}$ are given by
\begin{align*}
    &J(X) \isdef V_{\rmf, x} (x_{\rmr,\ell_\rmh}, \overline{x}_{\ell_\rmh}) + \sum_{i = 1}^{\ell_\rmh - 1} V_x (x_{\rmr,k}, \overline{x}_i) + \sum_{i = 0}^{\ell_\rmh-1}V_u (\overline{u}_i),\\
    &g_{\rm eq} (X) \isdef \matl \overline{x}_1 - f(x_k, \overline{u}_0) \\ \overline{x}_2 - f(\overline{x}_1, \overline{u}_1) \\ \vdots \\ \overline{x}_{\ell_\rmh} - f(\overline{x}_{\ell_\rmh - 1}, \overline{u}_{\ell_\rmh - 1})) \matr,\\
    &g_{\rm ineq} (X) \isdef \matl I_{\ell_\rmh \ell_u} & 0_{\ell_\rmh \ell_u \times \ell_\rmh \ell_x} \\ -I_{\ell_\rmh \ell_u} & 0_{\ell_\rmh \ell_u \times \ell_\rmh \ell_x} \matr X - \matl \mathds{1}_{\ell_\rmh \times 1} \otimes u_{\max} \\ -\mathds{1}_{\ell_\rmh \times 1} \otimes u_{\min} \matr.
\end{align*}
Then, the control input at step $k$ is calculated as shown in \eqref{eq:uk_NMPC}.
Instead of solving the general NLP given by \eqref{eq:NLP_J}$-$\eqref{eq:NLP_inequal}, which can be difficult and time-consuming, sequential quadratic programming (SQP) \cite{boggs1995,fernandez2010,izmailov2012} can be used to break down the NLP into subproblems given by quadratic programs (QP), which allow the solution to be updated iteratively until a stopping conditions is reached.
An example of an implementation of SQP to solve a NMPC problem based on CasADi \cite{andersson2019} is shown in \cite{cart_pendulum_MPC}, in which an inverted pendulum on a cart is stabilized at the upwards position;
the simulation in \cite{cart_pendulum_MPC} is used in Section \ref{sec:numerical_examples} to obtain data to train F-ARMA.

\subsection{Linear Model Predictive Control}\label{subsec:LMPC}
A simpler MPC formulation can be obtained in the case where, for all $k \ge 0,$ the dynamics of the discrete-time system are given by
\begin{align}
    x_{k+1} &= A x_k + B u_k, \label{eq:LD_x} \\
    y_k &= C x_k, \label{eq:LD_y}
\end{align}
where $A \in \BBR^{\ell_x \times \ell_x}$ is the state matrix, $B \in \BBR^{\ell_x \times \ell_u}$ is the input matrix, and $C \in \BBR^{\ell_y \times \ell_X}$ is the output matrix.
As in the NMPC case, it is assumed that, for all $k \ge 0,$ $x_k$ and a state reference $x_{\rmr,k} \in \BBR^{\ell_x}$ can be obtained from $y_k$ and $r_k.$
Then, the linear MPC problem to solve at each step $k > 0$ to minimize $\sum_{k = 0}^{\infty} V(r_k, y_k) = \sum_{k = 0}^{\infty} \Vert r_k - y_k \Vert$ is given by
\begin{align}
    &X_{{\rm mpc}, k} = \argmin_{X \in \BBR^{\ell_\rmh (\ell_x + \ell_u)}}  (x_{\rmr,k} - \overline{x}_{\ell_\rmh})^\rmT \overline{Q}_\rmf (x_{\rmr,k} - \overline{x}_{\ell_\rmh}) \nn \\
    & \hspace{2em} + \sum_{i = 1}^{\ell_\rmh-1} (x_{\rmr,k} - \overline{x}_i)^\rmT \overline{Q} (x_{\rmr,k} - \overline{x}_i) + \sum_{i = 0}^{\ell_\rmh-1} \overline{u}_i^\rmT \overline{R} \overline{u}_i, \label{eq:MPC_J}
\end{align}
subject to
\begin{align}
    &\overline{x}_1 = A x_k + B \overline{u}_0, \label{eq:MPC_equal_0}\\
    &\overline{x}_{i + 1} = A \overline{x}_i + B \overline{u}_i \mbox{ for all } i \in \{1, \ldots, \ell_\rmh-1 \}, \label{eq:MPC_equal}\\
    &\matl I_{\ell_u} \\ -I_{\ell_u} \matr \overline{u}_{i} \leq \matl u_{\rm max} \\ u_{\rm min} \matr \mbox{ for all } i \in \{0, \ldots, \ell_\rmh-1 \}, \label{eq:MPC_inequal}
\end{align}
where $\overline{Q}_\rmf \in \BBR^{\ell_x \times \ell_x}$ is the final state cost matrix, $\overline{Q} \in \BBR^{\ell_x \times \ell_x}$ is the state cost matrix and $\overline{R} \in \BBR^{\ell_u \times \ell_u}$ is the input cost matrix.
It is assumed that $\overline{Q}_\rmf$ and $\overline{Q}$ are symmetric, positive semidefinite matrices and that $\overline{R}$ is a symmetric, positive definite matrix.
Note that, for all $k > 0,$ it follows from \eqref{eq:LD_x} that
\begin{equation}
    x_k = A^k x_0 + \sum_{i = 0}^{k-1} A^{k-1-i} B u_i.\label{eq:x_k}
\end{equation}
Then, it follows from \eqref{eq:x_k} that \eqref{eq:MPC_equal_0},\eqref{eq:MPC_equal} can be written as
\begin{equation}
\matl \overline{x}_1 \\ \overline{x}_2 \\ \vdots \\ \overline{x}_{\ell_\rmh} \matr = \Gamma_\rmu \matl \overline{u}_0 \\ \overline{u}_1 \\ \vdots \\ \overline{u}_{\ell_\rmh - 1} \matr + \Gamma_\rmx x_k, \label{eq:xbar_k}
\end{equation}
where $\Gamma_\rmu \in \BBR^{\ell_\rmh \ell_x \times \ell_\rmh \ell_u}$ and $\Gamma_\rmx \in \BBR^{\ell_\rmh \ell_x \times \ell_x}$ are given by
\small
\begin{equation*}
\setlength\arraycolsep{2pt}
\Gamma_\rmu \isdef \matl 
B &  0_{\ell_x \times \ell_u} & \cdots & \cdots & \cdots & \cdots & 0_{\ell_x \times \ell_u} \\
AB &  B & 0_{\ell_x \times \ell_u} & \cdots & \cdots & \cdots & 0_{\ell_x \times \ell_u} \\
A^2 B & AB &  B & 0_{\ell_x \times \ell_u} & \cdots & \cdots & 0_{\ell_x \times \ell_u} \\
A^3 B & A^2 B &  AB & \ddots & \ddots & \ddots & \vdots \\
\vdots & \vdots & \vdots & \ddots & \ddots & \ddots & 0_{\ell_x \times \ell_u} \\
A^{\ell_\rmh - 1} B & A^{\ell_\rmh - 2} B & A^{\ell_\rmh - 3} B & \cdots & A^2 B & AB & B \\
\matr,
\end{equation*}
\normalsize
\begin{equation*}
\ \Gamma_\rmx \isdef \matl A^\rmT & (A^2)^\rmT & \cdots & (A^{\ell_\rmh})^\rmT \matr^\rmT.
\end{equation*}
Hence, it follows from \eqref{eq:xbar_k} that \eqref{eq:MPC_J}$-$\eqref{eq:MPC_inequal} can be written as the QP
\begin{equation}\label{eq:MPC_J_QP}
    U_{{\rm mpc}, k} = \argmin_{U \in \BBR^{\ell_\rmh \ell_u}} \frac{1}{2} U^\rmT H_{\rm mpc} U + q_{{\rm mpc},k}^\rmT U,
\end{equation}
subject to
\begin{equation}\label{eq:MPC_inequal_QP}
    \Gamma_{\rm mpc} \ U \le \nu_{\rm mpc},
\end{equation}
where
\begin{align*}
    H_{\rm mpc} &\isdef 2 \Gamma_\rmu^\rmT Q_{\rm mpc} \Gamma_\rmu + R_{\rm mpc},\\
    q_{{\rm mpc}, k} &\isdef -2 \Gamma_\rmu^\rmT Q_{\rm mpc} (\mathds{1}_{\ell_\rmh \times 1} \otimes x_{\rmr, k} - \Gamma_\rmx x_k),\\
    Q_{\rm mpc} &\isdef \matl I_{\ell_\rmh - 1} \otimes \overline{Q} & 0_{ (\ell_\rmh - 1) \ell_x \times \ell_x} \\ 0_{\ell_x \times (\ell_\rmh - 1) \ell_x} & \overline{Q}_\rmf  \matr, \\
    R_{\rm mpc} &\isdef I_{\ell_\rmh} \otimes \overline{R}, \\
    \Gamma_{\rm mpc} &\isdef \matl I_{\ell_\rmh \ell_u} \\ -I_{\ell_\rmh \ell_u} \matr \in \BBR^{2 \ell_\rmh \ell_u \times \ell_\rmh \ell_u},\\
    \nu_{\rm mpc} &\isdef \matl \mathds{1}_{\ell_\rmh \times 1} \otimes u_{\max} \\ -\mathds{1}_{\ell_\rmh \times 1} \otimes u_{\min} \matr \in \BBR^{2 \ell_\rmh \ell_u},\\
    U &\isdef \matl \overline{u}_0^\rmT & \cdots & \overline{u}_{\ell_\rmh -1}^\rmT \matr^\rmT \in \BBR^{\ell_\rmh \ell_u},\\
    U_{{\rm mpc}, k} &\isdef \matl u_{\rms, 0}^\rmT & \cdots & \overline{u}_{\rms, \ell_\rmh -1}^\rmT \matr^\rmT \in \BBR^{\ell_\rmh \ell_u}.
\end{align*}
Then, the control input at step $k$ is given by the initial control input solution, such that
\begin{equation}
    u_k = \overline{u}_{\rms, 0} = \matl I_{\ell_u} & 0_{\ell_u \times (\ell_\rmh - 1) \ell_u} \matr U_{{\rm mpc}, k}.\label{eq:uk_MPC}
\end{equation}
The QP given by \eqref{eq:MPC_J_QP},\eqref{eq:MPC_inequal_QP} can be solved by using several available QP solvers, including \texttt{quadprog} in Matlab.
This method will be used in used in Section \ref{sec:numerical_examples} to obtain data to train F-ARMA.


\section{Fuzzy-Autoregressive Moving Average Controller} \label{sec:F-ARMA}

This section introduces the Fuzzy-Autoregressive Moving Average (F-ARMA) control technique, which is motivated by the Takagi-Sugeno fuzzy systems \cite[ch.~6]{lilly2011}.
%
%

\subsection{ARMA controllers}

The F-ARMA controller consists of $\ell_\rmc>0$ ARMA controllers. 
Each ARMA controller is trained to emulate the response of the MPC controller under particular conditions.
The $i$th ARMA controller's request control, denoted by 
$u_{\rmr, i, k},$ is given by
\begin{equation}\label{eq:ARMA_0}
    u_{\rmr, i, k} = 
    \begin{cases}
    0, & k < \ell_{\rmw, i}, \\
    \sum_{j = 1}^{\ell_{\rmw, i}} D_{i,j} u_{i, k - j} + \sum_{j = 1}^{\ell_{\rmw, i}} N_{i,j} z_{i, k - j}, & k \ge \ell_{\rmw, i},
    \end{cases}
\end{equation}
where $\ell_{\rmw, i} > 0$ is the window length of the ARMA controller, $D_{i,j} \in \BBR^{\ell_u \times \ell_u}$ and $N_{i,j} \in \BBR^{\ell_u \times \ell_y}$ are controller coefficient matrices, $u_{i, k} \isdef \sigma (u_{\rmr, i, k}) \in \BBR^{\ell_u}$ is the constrained control input, and $z_{i, k} \in \BBR^{\ell_y}$ is the performance variable of the $i$th ARMA controller. 

To simplify notation, \eqref{eq:ARMA_0} can be written as
\begin{equation}\label{eq:ARMA}
    u_{\rmr, i, k} = 
    \begin{cases}
    0, & k < \ell_{\rmw, i}, \\
    \phi_{i,k} \theta_i, & k \ge \ell_{\rmw, i},
    \end{cases}
\end{equation}
where the regressor matrix $\phi_{i,k} \in \BBR^{\ell_u \times \ell_{\theta,i}}$ is given by
\begin{equation*}
    \setlength\arraycolsep{2pt}
    \phi_{i,k} 
        \isdef
            I_{\ell_u} \otimes
            \matl u_{i, k-1}^\rmT & \cdots & u_{i, k-\ell_{\rmw, i}}^\rmT & z_{i, k-1}^\rmT & \cdots & z_{i, k-\ell_{\rmw, i}}^\rmT \matr ,
\end{equation*}
the controller coefficient vector $\theta_i \in \BBR^{\ell_{\theta,i}}$ is given by
\begin{equation*}
    \theta_i \isdef {\rm vec} \matl D_{i,1} & \cdots & D_{i, \ell_{\rmw, i}} & N_{i,1} \cdots N_{i, \ell_{\rmw, i}} \matr^\rmT,
\end{equation*}
and $\ell_{\theta,i} \isdef \ell_{\rmw, i} \ell_u (\ell_y + \ell_u).$

The performance variable of the $i$the ARMA controller 
$z_{i, k} \isdef \SZ_i(r_k, y_k) \in \BBR^{\ell_y},$ 
where $\SZ_i \colon \BBR^{\ell_y} \times \BBR^{\ell_y} \to \BBR^{\ell_y}$ calculates the performance variable to meet the same tracking objective as $V$ within the neighborhood in which the $i$-th ARMA controller is more favored by the F-ARMA controller to better emulate the response of the original MPC controller;
in most cases, $\SZ_i(r_k, y_k) = r_k - y_k$ for all $k\ge 0.$
Note that, unlike the MPC techniques shown in Section \ref{sec:MPC}, the F-ARMA controllers use $y_k$ and $r_k$ as inputs, rather than $x_k$ and $x_{\rmr,k},$ and thus would not require an observer for implementation.

\subsection{Fuzzy-ARMA Controller}
Consider and T-S fuzzy system with $\ell_\rmc$ rules associated with each of the ARMA controllers, let $\gamma_k \in \BBR^{\ell_\gamma}$ the vector of decision variables chosen to determine the favored controller at each step $k\ge0,$ and, for all $j \in \{1, \ldots, \ell_\gamma\}$ let $\gamma_{k,j} \in \BBR$ be the $j$th component of $\gamma_k,$ such that $\gamma_k = \matl \gamma_{k,1} & \cdots & \gamma_{k,\ell_\gamma} \matr^\rmT.$
Then, rule $\SR_i$ can be written as
\begin{align}
    \SR_i \colon \quad &\mbox{If } \gamma_{k,1} \mbox{ is } \SF_{i, 1} \mbox{ and } \gamma_{k,2} \mbox{ is } \SF_{i, 2} \mbox{ and } \cdots \nn \\
    &\mbox{and } \gamma_{k,\ell_\gamma} \mbox{ is } \SF_{i, \ell_\gamma}, \mbox{ then } u_{\rmr, i, k} = \phi_{i,k} \theta_i,
\end{align}
where, for all $j \in \{1, \ldots, \ell_\gamma\},$ $\SF_{i, j}$ is a fuzzy set characterized by the membership function $\mu_{i,j} \colon \BBR \to \BBR,$ which quantifies how much $\gamma_{k,j}$ belongs to $\SF_{i, j}.$

Next, considering a product T-norm as an implication method, let $w_{i, k} > 0$ be the weight associated with rule $\SR_i$ at step $k$ and be given by
\begin{equation}
    w_{i, k} \isdef \prod_{j = 1}^{\ell_\gamma} \mu_{i,j}(\gamma_{k,j}). \label{eq:wik}
\end{equation}
Then, the requested control input generated by F-ARMA is given by the weighted average of the constrained control inputs generated by each ARMA controller, that is, 
\begin{equation}
    u_{\rmr, k} = \frac{\sum_{i = 1}^{\ell_\rmc} w_{i, k} u_{i, k}}{\sum_{i = 1}^{\ell_\rmc} w_{i, k}} = \frac{\sum_{i = 1}^{\ell_\rmc} w_{i, k} \sigma(\phi_{i,k} \theta_i)}{\sum_{i = 1}^{\ell_\rmc} w_{i, k}}.\label{eq:urk}
\end{equation}
A method to obtain $\theta_i$ from data samples acquired from MPC closed-loop simulations is introduced in Section \ref{sec:LS}. 
In this work, the vector $\gamma_k,$ the fuzzy rules $\SR_i,$ and the membership functions $\mu_{i,j}$ are defined arbitrarily by the user to better interpolate the linear behaviors of each ARMA controller.


\section{Least-Squares Regression for ARMA Controller Synthesis} \label{sec:LS}

Let $i \in \{1, \ldots, \ell_\rmc\},$ and $\ell_{{\rm tr},i} > 0$ be the number of data samples used for the training the $i$th ARMA controller.
For all $k \in [0, \ell_{{\rm tr},i} - 1],$ let $u_{{\rm tr}, i, k} \in \BBR^{\ell_u}$ and $y_{{\rm tr}, i, k} \in \BBR^{\ell_y}$ be the input and output data obtained from the implementation of MPC for closed-loop trajectory tracking, and let $r_{{\rm tr}, i,k} \in \BBR^{\ell_y}$ be the chosen reference to train the $i$th ARMA controller; 
furthermore, let $z_{{\rm tr}, i,k} \in \BBR^{\ell_y}$ be the training performance variables given by $z_{{\rm tr}, i,k} = \SZ_i (r_{{\rm tr}, i,k}, y_{{\rm tr}, i,k}).$
Then, $\theta_i$ can be obtained by solving the least squares problem
\begin{equation} \label{eq:LS}
    \theta_i = \argmin_{\theta \in \BBR^{\ell_{\theta, i}}} \Vert \Phi_{{\rm tr}, i} \theta - U_{{\rm tr}, i} \Vert + 
     \theta^\rmT R_{\theta,i} \theta,
\end{equation}
where $R_{\theta,i} \in \BBR^{\ell_{\theta,i} \times \ell_{\theta,i}}$ is a positive semidefinite, symmetric regularization matrix,
%
%
\footnotesize
\begin{equation*}
    \setlength\arraycolsep{2pt}
    \Phi_{{\rm tr}, i} \isdef \matl \phi_{{\rm tr}, i,\ell_{\rmw, i}}^\rmT & \cdots & \phi_{{\rm tr}, i, \ell_{{\rm tr}, i}-1}^\rmT \matr^\rmT \in \BBR^{\ell_u (\ell_{{\rm tr}, i} - \ell_{\rmw, i} - 1) \times \ell_{\theta,i}},
\end{equation*}
\begin{equation*}
    \setlength\arraycolsep{2pt}
    \phi_{{\rm tr}, i,k} \isdef I_{\ell_u} \otimes \matl u_{{\rm tr}, i, k-1}^\rmT & \cdots & u_{{\rm tr}, i, k-\ell_{\rmw, i}}^\rmT & z_{{\rm tr}, i, k-1}^\rmT & \cdots & z_{{\rm tr}, i, k-\ell_{\rmw, i}}^\rmT \matr 
\end{equation*}
\begin{equation*}
    \hspace{-19em}\in \BBR^{\ell_u \times \ell_{\theta,i}},
\end{equation*}
\begin{equation*}
    \setlength\arraycolsep{2pt}
    U_{{\rm tr}, i} \isdef \matl u_{{\rm tr}, i, \ell_{\rmw, i}}^\rmT & \cdots & u_{{\rm tr}, i, \ell_{{\rm tr}, i}-1}^\rmT \matr^\rmT \in \BBR^{\ell_u (\ell_{{\rm tr}, i} - \ell_{\rmw, i} - 1)}.
\end{equation*}
\normalsize
While linear controllers cannot enforce input constraints, \eqref{eq:LS} can be formulated as a QP to obtain a controller that more easily meets the constraints shown in \eqref{eq:Hsat1},\eqref{eq:Hsat2}.
Hence, \eqref{eq:LS} with input constraints can be reformulated as the following QP
\begin{equation}\label{eq:QP_1}
     \theta_i = \argmin_{\theta \in \BBR^{\ell_{\theta, i}}} \frac{1}{2} \theta^\rmT H_{{\rm tr},i} \theta + q_{{\rm tr},i}^\rmT \theta,
\end{equation}
subject to
\begin{equation}\label{eq:QP_2}
    \Gamma_{{\rm tr},i} \theta \le \nu_{{\rm tr},i},
\end{equation}
where
\begin{align*}
    H_{{\rm tr},i} &\isdef 2 \Phi_{{\rm tr}, i}^\rmT \Phi_{{\rm tr}, i} + 2 R_{\theta,i} \in \BBR^{\ell_{\theta,i} \times \ell_{\theta,i}}, \\
    q_{{\rm tr},i} &\isdef - 2 \Phi_{{\rm tr}, i}^\rmT U_{{\rm tr}, i} \in \BBR^{\ell_{\theta,i}}, \\
    \Gamma_{{\rm tr},i} &\isdef \matl \Phi_{{\rm tr}, i} \\ -\Phi_{{\rm tr}, i} \matr \in \BBR^{2 \ell_u (\ell_{{\rm tr}, i} - \ell_{\rmw, i} - 1) \times \ell_{\theta,i}}, \\
    \nu_{{\rm tr},i} &\isdef \matl \mathds{1}_{(\ell_{{\rm tr}, i} - \ell_{\rmw, i} - 1)\times 1} \otimes u_{\max} \\ -\mathds{1}_{(\ell_{{\rm tr}, i} - \ell_{\rmw, i} - 1)\times 1} \otimes u_{\min} \matr \in \BBR^{2 \ell_u (\ell_{{\rm tr}, i} - \ell_{\rmw, i} - 1)}.
\end{align*}
The QP given by \eqref{eq:QP_1}, \eqref{eq:QP_2} can be solved in Matlab using \texttt{quadprog} to obtain $\theta_i$ corresponding to the $i$th ARMA controller of F-ARMA.


\section{Numerical Examples} \label{sec:numerical_examples}

In this section, we implement MPC algorithms presented in Section \ref{sec:MPC} to obtain data to train F-ARMA controllers presented in Section \ref{sec:F-ARMA} and demonstrate its performance at emulating the response of the original MPC algorithms.
The continuous-time dynamics in the examples are simulated in Matlab by using \texttt{ode45} every $T_\rms$ s, such that $u$ is kept constant in each simulation interval.
In Example \ref{ex:double_integrator}, the objective is the setpoint tracking of a double integrator system with input constraints, in which linear MPC is implemented to obtain data.
In Example \ref{ex:cart_pendulum}, the objective is to invert a pendulum on a cart with horizontal force constraints, in which NMPC is implemented to obtain the training data.

\begin{exam}\label{ex:double_integrator}
{\textbf{\textit{Setpoint tracking of double integrator system with input constraints.}}}
Consider the double integrator
\begin{align}
\dot{x} &= \matl 0 & 1 \\ 0 & 0 \matr x + \matl 0 \\ 1 \matr u, \label{eq:xeq_ex1} \\
y &= \matl 1 & 0 \matr x, \label{eq:yeq_ex1}
\end{align}
where $x \in \BBR^2,$ $u \in \BBR,$ and $y \in \BBR.$
In this example, the objective is for the system output to reach a value of 2, such that $\lim_{t \to \infty} y(t) = 2$ under the input constraints $u_{\rm max} = -u_{\rm min} = 10.$
%
Hence, for all $k\ge0,$ the reference signals are given by $r_k = 2,$ and $x_{\rmr, k} = \matl 2 & 0 \matr^\rmT.$

Using exact discretization of \eqref{eq:xeq_ex1},\eqref{eq:yeq_ex1} with sampling time $T_\rms > 0$ yields the discrete-time dynamics
\begin{align}
x_{k+1} &= \matl 1 & T_\rms \\ 0 & 1 \matr x_k + \matl T_\rms^2 \\ T_\rms \matr u_k = A x_k + B u_k, \label{eq:xdteq_ex1} \\
y_k &= \matl 1 & 0 \matr x_k = C x_k \label{eq:ydteq_ex1}
\end{align}
for all $k\ge0,$ where $x_k \in \BBR^2,$ $u_k \in \BBR,$ and $y_k \in \BBR.$
In this example, we set sampling rate $T_\rms = 0.01$ s.
The linear MPC algorithm introduced in Subsection \ref{subsec:LMPC} is applied to \eqref{eq:xeq_ex1},\eqref{eq:yeq_ex1} to obtain data to train the F-ARMA controllers, and designed by considering the discrete-time dynamics \eqref{eq:xdteq_ex1},\eqref{eq:ydteq_ex1} with $\ell_\rmh = 10,$ $\overline{Q} = 10 I_2,$ $\overline{Q}_\rmf = 10^5 I_2,$ and $\overline{R} = 10^{-2} I_2.$ 
The results from implementing linear MPC on the double integrator dynamics for $x(0) = \matl 0 & 0\matr^\rmT$ are shown in Figure \ref{fig:ex1_train}.
The $y$ and $u$ data from this simulation is used to train the linear controllers of F-ARMA.
\begin{figure}[h!]
\vspace{1em}
    \centering
    \includegraphics[width=0.9\columnwidth]{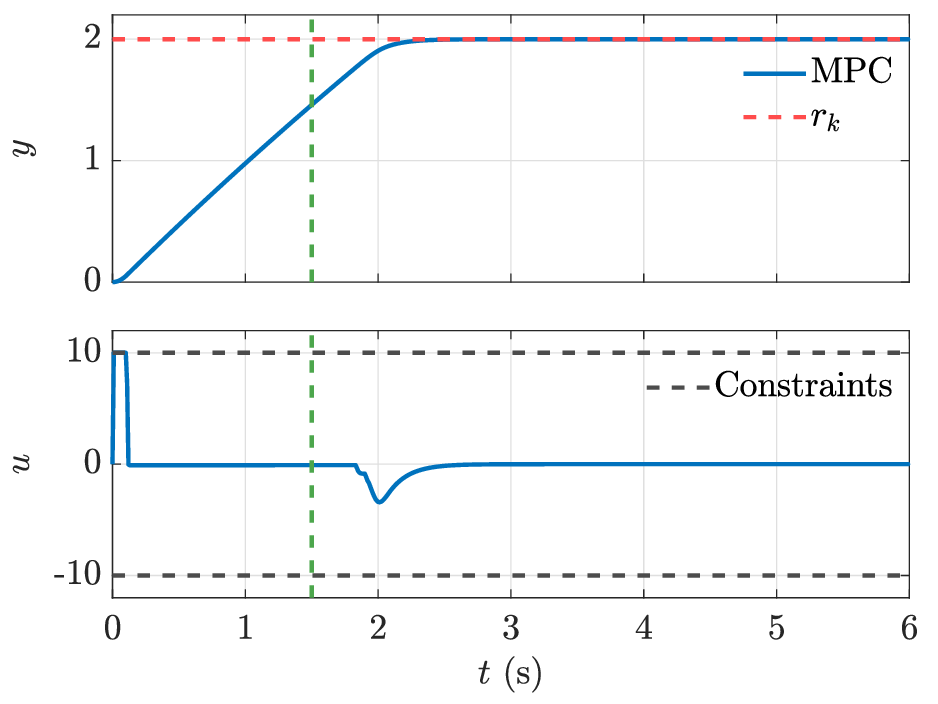}
    \vspace{-0.5em}
    \caption{Example \ref{ex:double_integrator}: {\bf Double Integrator}. Results from implementing linear MPC on the double integrator dynamics given by \eqref{eq:xeq_ex1},\eqref{eq:yeq_ex1} for $x(0) = \matl 0 & 0\matr^\rmT$.
    Controller coefficient vector $\theta_1$ corresponding to the first F-ARMA linear controller is trained using the $y$ and $u$ data corresponding to $t \in [0, 6]$ s, that is, all the data shown in the figure.
    Controller coefficient vector $\theta_2$ corresponding to the second F-ARMA linear controller is trained using the $y$ and $u$ data corresponding to $t \in [1.5, 6]$ s, that is, all the data from $t = 1.5$ s, denoted by the vertical, dashed green line, onwards.
    } 
    \label{fig:ex1_train}
     \vspace{-1em}
\end{figure}

The F-ARMA implementation for this example considers two linear controllers, that is,  $\ell_\rmc = 2.$
The first one is favored at the steps $k\ge0$ in which $|r_k - y_k|$ is large, whereas 
the second one is favored at the steps $k\ge0$ in which $|r_k - y_k|$ is small.
For this purpose, for all $k \ge 0,$ we consider the decision variable $\gamma_k = \gamma_{k, 1} = |r_k - y_k|,$ and the fuzzy sets $\SF_{1,1}$ and $\SF_{2,1}$ corresponding to the cases where $\gamma_k$ is large and small, respectively.
Hence, the set of fuzzy rules can be written as
\begin{align*}
    \SR_1 \colon \quad &\mbox{If } \gamma_k \mbox{ is } \mbox{Large} \mbox{ then } u_{\rmr, 1, k} = \phi_{1,k} \theta_1,\\
    \SR_2 \colon \quad &\mbox{If } \gamma_k \mbox{ is } \mbox{Small} \mbox{ then } u_{\rmr, 2, k} = \phi_{2,k} \theta_2.
\end{align*}
For simplicity, let $\mu_1 = \mu_{1,1}$ and $\mu_2 = \mu_{2,1}$ denote the membership functions associated with fuzzy sets $\SF_{1,1}$ (large) and $\SF_{2,1}$ (small), respectively, and be given by
\begin{equation*}
    \mu_1 (\gamma_k) = \begin{cases}
        0, & \gamma_k < 0.4,\\
        1, & \gamma_k >0.6,\\
        (\gamma_k - 0.4) / 0.2,& \mbox{otherwise},
    \end{cases}
\end{equation*}
\begin{equation*}
    \mu_2 (\gamma_k) = \begin{cases}
        1, & \gamma_k < 0.4,\\
        0, & \gamma_k >0.6,\\
        (-\gamma_k + 0.6) / 0.2,& \mbox{otherwise},
    \end{cases}
\end{equation*}
as shown in Figure \ref{fig:ex1_mu_fuzzy}.
Hence, for all $k \ge 0,$ it follows from \eqref{eq:wik},\eqref{eq:urk} that the control input calculated by the F-ARMA controller is given by
\begin{equation}
    u_{\rmr, k} = \frac{\mu_1 (\gamma_k) \ \sigma(\phi_{1,k} \theta_1) + \mu_2 (\gamma_k) \ \sigma(\phi_{2,k} \theta_2)}{\mu_1 (\gamma_k) + \mu_2 (\gamma_k)}.
\end{equation}
The ARMA controller coefficient vectors $\theta_1$ and $\theta_2$ are obtained by solving \eqref{eq:QP_1}, \eqref{eq:QP_2} with $\ell_{\rmw,1} = \ell_{\rmw,2} = 10,$ $\SZ_1(r_k, y_k) = \SZ_2(r_k, y_k) = r_k - y_k$ for all $k \ge 0,$ and $R_{\theta,1} = R_{\theta,2} = 0.$
For $\theta_1,$ the $y$ and $u$ data obtained from the linear MPC simulation results shown in Figure \ref{fig:ex1_train} corresponding to $t \in [0, 6]$ s is used for training, such that $\ell_{{\rm tr}, 1} = 601$ and, for all $k \in \{0, \ldots, \ell_{{\rm tr}, 1}-1\},$ $u_{{\rm tr}, 1, k} = u_k$ and $y_{{\rm tr}, 1, k} = y_k.$
For $\theta_2,$ the $y$ and $u$ data obtained from the linear MPC simulation results shown in Figure \ref{fig:ex1_train} corresponding to $t \in [1.5, 6]$ s is used for training, such that $\ell_{{\rm tr}, 2} = 451$ and, for all $k \in \{0, \ldots, \ell_{{\rm tr}, 2}-1\},$ $u_{{\rm tr}, 2, k} = u_{150 + k}$ and $y_{{\rm tr}, 2, k} = y_{150 + k}.$
\begin{figure}[h!]
    \vspace{-2em}
    \centering
    \includegraphics[width=0.9\columnwidth]{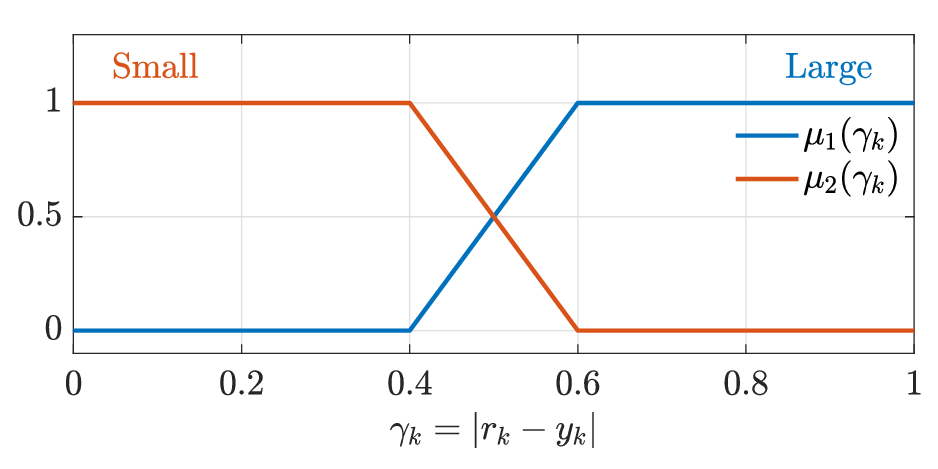}
    \vspace{-0.5em}
    \caption{Example \ref{ex:double_integrator}: {\bf Double Integrator}. Fuzzy membership functions $\mu_1$ and $\mu_2$ corresponding to the fuzzy sets in which $\gamma_k$ is large and small, respectively.} 
    \label{fig:ex1_mu_fuzzy}
     \vspace{-1em}
\end{figure}

The results from implementing the trained F-ARMA controller on the double integrator dynamics for $x(0) = \matl 0 & 0\matr^\rmT$ are shown in Figure \ref{fig:ex1_yu}.
The response of F-ARMA is compared against the response of linear MPC and the individually trained ARMA controllers with $\theta_1$ and $\theta_2$ implemented using \eqref{eq:ARMA}.
While the individual ARMA controllers are able to reach the desired setpoint, the transient response of the ARMA controller with $\theta_1$ is much slower than the response of MPC, leading to the slower convergence shown in the $y$ versus $t$ plot, and the response of the ARMA controller with $\theta_2$ is more aggressive than that of MPC, leading to the increased control effort, as shown in the $u$ versus $t$ plot.
\begin{figure}[h!]
    \centering
    \vspace{-0.75em}
    \includegraphics[width=0.9\columnwidth]{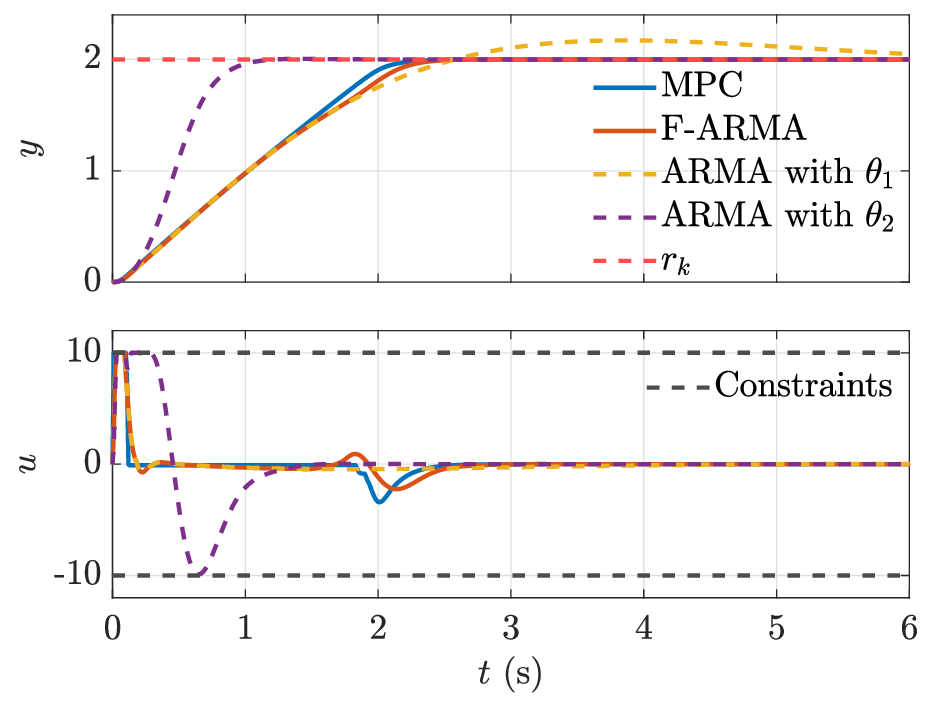}
    \vspace{-0.5em}
    \caption{Example \ref{ex:double_integrator}: {\bf Double Integrator}. Results from implementing the trained F-ARMA controller on the double integrator dynamics given by \eqref{eq:xeq_ex1},\eqref{eq:yeq_ex1} for $x(0) = \matl 0 & 0\matr^\rmT$.
    The response of F-ARMA is compared against the response of linear MPC and the individually trained ARMA controllers with $\theta_1$ and $\theta_2$ implemented using \eqref{eq:ARMA}.} 
    \label{fig:ex1_yu}
\end{figure}

Unlike the individual ARMA controllers, the F-ARMA controller closely approximates the response of the original linear MPC controller.
Note that the input constraints are maintained without explicitly applying constraints.
Furthermore, the average times taken to run linear MPC and F-ARMA at each step are $1.1 \times 10^{-3}$ s and $8.2 \times 10^{-6}$ s, respectively.
\hfill {\LARGE$\diamond$}
\end{exam}

\begin{exam}\label{ex:cart_pendulum}
\textbf{\textit{Swing up of inverted pendulum on a cart with horizontal force constraints.}}
Consider the inverted pendulum on a cart system shown in Figure \ref{fig:cart_pendulum}, in which a pendulum is attached to a cart that moves horizontally at a frictionless pivot point.
Let $p \in \BBR$ be the horizontal position relative to a reference point, $\phi \in \BBR$ be the angle of the pendulum relative to its upwards position, let $F \in \BBR$ be the horizontal force applied to the cart, and let the acceleration due to gravity $g$ have a downwards direction.
Defining the state vector $x \isdef \matl p & \dot{p} & \phi & \dot{\phi} \matr^\rmT \in \BBR^4$ and the input $u \isdef F \in \BBR,$ the dynamics of the inverted pendulum on a cart is given by
\begin{align}
\dot{x} & = f_\rmc \left(\matl p & \dot{p} & \phi & \dot{\phi} \matr^\rmT, F \right) \nn \\ 
&\isdef\matl \dot{p} \\
\frac{ \tfrac{1}{6} m^2 \ell^3 \dot{\phi}^2 \sin\phi \ - \ \tfrac{1}{8} (m\ell)^2 g \sin 2\phi \ + \ \tfrac{1}{3} m\ell^2 F}{\tfrac{1}{3} m\ell^2 (m \ + \ M) \ - \ \tfrac{1}{4} (m \ell \cos\phi)^2} \\ 
\dot{\phi} \\ 
\frac{ \tfrac{1}{2} mg\ell(m \ + \ M) \sin\phi \ - \ \tfrac{1}{8} (m\ell)^2 \dot\phi^2 \sin 2\phi \ - \ \tfrac{1}{2} (m \ell \cos\phi) F }{\tfrac{1}{3} m\ell^2 (m \ + \ M) \ - \ \tfrac{1}{4} (m \ell \cos\phi)^2} \matr, \label{eq:xeq_ex2}\\
y &= \matl p \\ \phi \matr = C x \isdef \matl 1 & 0 & 0 & 0 \\ 0 & 0 & 1 & 0 \matr x. \label{eq:yeq_ex2}
\end{align}
%
where $f_\rmc \colon \BBR^4 \times \BBR \to \BBR^4,$ $M > 0$ is the mass of the cart, $m > 0$ is the mass of the pendulum, and $\ell > 0$ is the length of the pendulum.

In this example, to simulate the pendulum on a cart system, we set $M =1$ kg, $m = 0.2$ kg, $L = 0.4$ m, $g = 9.81$ m/s$^2.$
The objective is to invert the pendulum from a downward orientation, 
that is, $\phi = \pi$, 
and stabilize the upward orientation under the input constraints $u_{\rm max} = -u_{\rm min} = 30,$ and sampling rate $T_\rms = 0.02$ s.
Note that $x = 0$ is an unstable equilibrium.
%
Hence, for all $k\ge0,$ the reference signals are given by $r_k = 0,$ and $x_{\rmr, k} = 0.$

\begin{figure}[ht]
\vspace{-1em}
\centering
\resizebox{0.4\columnwidth}{!}{
\begin{tikzpicture}[>={stealth'}, line width = 0.25mm] 
%
\node[draw = none] at (0,0) (orig) {}; 
\node[draw = none] at ([xshift = 1.8 em]orig.center) (orig_w) {}; 
\node[draw = none] at ([xshift = 2.5 em, yshift = 7 em]orig.center) (mass){}; 
\node[draw = none] at ([yshift = 7 em]orig.center) (ver_p){}; 

%
\node[smallblock,  rounded corners, minimum height = 2em, minimum width = 3em] at (orig.center) (P_point) {}; 
\node[smallblock, fill = black!20, rounded corners,minimum height = 8.25em, minimum width = 0.75em,rotate around={-19.7:(orig.center)}] at ([yshift = 3.7em]orig.center) (rod) {};
%
\draw[red,thick,dashed] (orig.center) -- (ver_p.center); 
\draw[->, red,thick, line width = 0.2mm] ([yshift = 3.15em]orig) arc (90:70:3em); 
\node[draw = none, red] at ([xshift = 0.45em, yshift = 3.75em]orig) {\footnotesize$\phi$}; 
\draw[red,thick,dashed] (orig.center) -- (mass.center); 
\node[draw, fill=black, circle, inner sep=0.075em] at (orig.center) {}; 
\node[draw = none] at ([xshift = -0.75em, yshift = 0.4em]orig.center) {\footnotesize$c$};
\node[draw = none] at ([xshift = 0.9em]mass.center) {\footnotesize$\SP$};
\node[draw = none] at ([xshift = 2em, yshift = 0.5em]orig.center) {\footnotesize$\SC$};

\draw[->, red]([xshift = -2em]P_point.west) -- node[above, yshift = -0.1em]{\footnotesize$F$}(P_point.west);
\node[draw, fill=black!20, circle, node distance = 1em, inner sep = 0.25em] at ([xshift = -0.8em, yshift = -1.35em]orig.center) {};
\node[draw, fill=black!20, circle, node distance = 1em, inner sep = 0.25em] at ([xshift = 0.8em, yshift = -1.35em]orig.center) {};
\node (grnd1) [ground,rotate=0, minimum width=7em, minimum height = 0.2em, inner sep = 0.2em] at ([xshift = -0.2em, yshift = -1.95em]P_point.west) {}; 
\draw (grnd1.north east) -- (grnd1.north west); 
\node[draw, fill=black, circle, inner sep=0.075em] at ([xshift = 1em]grnd1.north west) {}; %
\node[draw = none] at ([xshift = 0.5em, yshift = 0.5em]grnd1.north west) {\footnotesize$w$};
\draw[->,red,thick,dashed] ([xshift = 1em, yshift = -1em]grnd1.north west) -- node[below,yshift = 0.1em]{\footnotesize$p$} ([xshift = 5.275em, yshift = -1em]grnd1.north west);
\draw[-,red,thick] ([xshift = 1em, yshift = -1.3em]grnd1.north west) -- ([xshift = 1em, yshift = -0.7em]grnd1.north west);
\draw[-,red,thick] ([xshift = 5.275em, yshift = -1.3em]grnd1.north west) -- ([xshift = 5.275em, yshift = -0.7em]grnd1.north west);

\draw[->, red] ([xshift = -1.5em, yshift = 6.5em]orig.center) -- node[xshift = 0.4em, yshift = 0.25em] {\footnotesize $g$} ([xshift = -1.5em, yshift = 5em]orig.center); 
\end{tikzpicture}
}
\vspace{-0.5em}
\caption{\footnotesize 
Example \ref{ex:cart_pendulum}:
{\bf Inverted Pendulum on Cart}.
The pendulum $\SP$ is attached to the cart $\SC$ at the pivot point $c,$ $p$ is the horizontal position from $c$ to a reference point in the ground $w,$ $\phi$ is the angle of $\SP$ relative to its upwards position, and the horizontal force $F$ is applied to $\SC.$
The acceleration due to gravity $g$ has a downwards direction.}
\label{fig:cart_pendulum}
\end{figure}
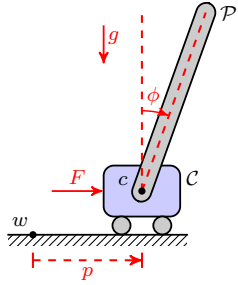

Using the Euler discretization of \eqref{eq:xeq_ex2},\eqref{eq:yeq_ex2} with sampling time $T_\rms > 0,$ the discrete-time dynamics is
\begin{align}
    x_{k+1} &= f(x_k, u_k) = x_k + T_\rms f_\rmc(x_k, u_k), \label{eq:xdteq_ex2}\\
    y_k &= h(x_k, u_k) = C x_k, \label{eq:ydteq_ex2}
\end{align}
for all $k\ge0,$ where $x_k \in \BBR^4,$ $u_k \in \BBR,$ and $y_k \in \BBR^2.$
The NMPC algorithm reviewed in Section \ref{sec:MPC} is applied to \eqref{eq:xeq_ex2},\eqref{eq:yeq_ex2} to obtain data to train the F-ARMA controllers, and designed by considering the discrete-time dynamics \eqref{eq:xdteq_ex2},\eqref{eq:ydteq_ex2},
with$\ell_\rmh = 100,$
%
\begin{align*}
&V_\rmf(\overline{x}_\rmr, \overline{x}) = V_x (\overline{x}_\rmr, \overline{x}) \\
&\hspace{0.25em} = \matl \overline{x}_{(1)} \\ \overline{x}_{(2)} \\ 1 - \cos(\overline{x}_{(3)}) \\ \overline{x}_{(4)} \matr^\rmT \matl 30 & 0 & 0 & 0 \\ 0 & 20 & 0 & 0 \\ 0 & 0 & 60 & 0 \\ 0 & 0 & 0 & 20 \matr \matl \overline{x}_{(1)} \\ \overline{x}_{(2)} \\ 1 - \cos(\overline{x}_{(3)}) \\ \overline{x}_{(4)} \matr
\end{align*}
%
for all $\overline{x}_\rmr, \overline{x} \in \BBR^4,$ and $V_u (\overline{u}) = 50 \overline{u}^2$ for all $\overline{u} \in \BBR.$
Note that the form of $V_\rmf$ and $V_x$ encode the objective of stabilizing the pendulum at its upwards position, that is, minimizing $1 - \cos \phi.$
The NMPC controller is implemented using SQP and CasADi, as shown in the \texttt{CartPendulum\_NonlinearMPC\_SwingUp.m} file in \cite{cart_pendulum_MPC}.
NMPC is implemented on the inverted pendulum on cart dynamics for two cases: a swing-up maneuver with $x(0) = \matl 0 & 0 & \pi & 0\matr^\rmT$ and a stabilization near the equilibrium maneuver with $x(0) = \matl 1 & 0 & \pi/5 & 0 \matr^\rmT.$
Each of these cases encode a different behavior that each F-ARMA linear controller needs to emulate.
The simulation results are shown in Figure \ref{fig:ex1_train}.
The $y$ and $u$ data from this simulation is used to train the linear controllers of F-ARMA.
Henceforth, for all $k \in \{0, 15/T_\rms\},$ the data obtained from the swing-up maneuver is denoted as $y_{{\rm su}, k},$ $u_{{\rm su}, k},$ and the data obtained from the stabilization near equilibrium maneuver is denoted as $y_{{\rm st}, k},$ $u_{{\rm st}, k}.$

\begin{figure}[h!]
    \vspace{-1em}
    \centering
    \includegraphics[width=0.85\columnwidth]{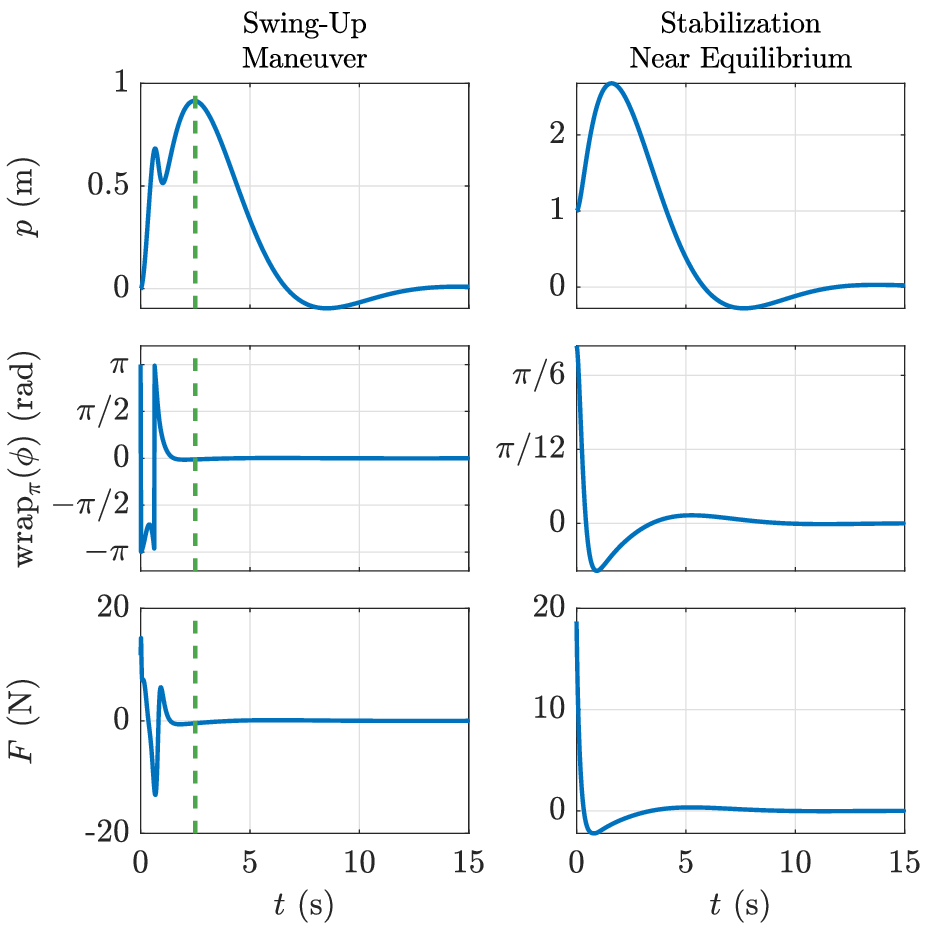}
    \vspace{-0.5em}
    \caption{Example \ref{ex:cart_pendulum}: {\bf Inverted Pendulum on Cart}. Results from implementing NMPC on the double integrator dynamics given by \eqref{eq:xeq_ex2},\eqref{eq:yeq_ex2} for a swing-up maneuver and stabilization near equilibrium, corresponding to the plots in the left and right columns, respectively. 
    Controller coefficient vector $\theta_1$ corresponding to the first F-ARMA linear controller is trained using the $y = \matl p & \phi \matr^\rmT$ and $u = F$ data from the swing-up maneuver simulation corresponding to $t \in [0, 2.5]$ shown in the left column plots before $t = 2.5$ s, denoted by the vertical, dashed green line.
    Controller coefficient vector $\theta_2$ corresponding to the second F-ARMA linear controller is trained using the $y = \matl p & \phi \matr^\rmT$ and $u = F$ data from the stabilization near equilibrium simulation corresponding to $t \in [0, 15]$ shown in the right column plots.} 
    \label{fig:ex2_train}
     \vspace{-1em}
\end{figure}

The F-ARMA implementation for this example considers two linear controllers, such that $\ell_\rmc = 2.$
The first one is favored at the steps $k\ge0$ in which the pendulum is not near the pendulum upwards orientation, whereas 
the second one is favored at the steps $k\ge0$ in which the pendulum is near the pendulum upwards orientation.
For this purpose, for all $k \ge 0,$ we consider the decision variable $\gamma_k = \gamma_{k, 1} = |\mbox{wrap}_\pi (\phi_k)| = \left|\mbox{wrap}_\pi \left(\matl 0 & 1 \matr y_k \right)\right|,$ and the fuzzy sets $\SF_{1,1}$ and $\SF_{2,1}$ corresponding to the cases where $\gamma_k$ is not near the upwards orientation ($-$ NUO) and near the upwards orientation (NUO), respectively.
Hence, the set of fuzzy rules can be written as
\begin{align*}
    \SR_1 \colon \quad &\mbox{If } \gamma_k \mbox{ is } - \mbox{NUO} \mbox{ then } u_{\rmr, 1, k} = \phi_{1,k} \theta_1,\\
    \SR_2 \colon \quad &\mbox{If } \gamma_k \mbox{ is } \mbox{NUO} \mbox{ then } u_{\rmr, 2, k} = \phi_{2,k} \theta_2.
\end{align*}
For simplicity, let $\mu_1 = \mu_{1,1}$ and $\mu_2 = \mu_{2,1}$ denote the membership functions associated with fuzzy sets $\SF_{1,1}$ ($-$ NUO) and $\SF_{2,1}$ (NUO), respectively, and be given by
\begin{equation*}
    \mu_1 (\gamma_k) = \begin{cases}
        0, & \gamma_k < \tfrac{\pi}{3} - \tfrac{\pi}{30},\\
        1, & \gamma_k > \tfrac{\pi}{3} + \tfrac{\pi}{30},\\
        (\gamma_k - (\tfrac{\pi}{3} - \tfrac{\pi}{30})) / (\tfrac{\pi}{15}),& \mbox{otherwise},
    \end{cases}
\end{equation*}
\begin{equation*}
    \mu_2 (\gamma_k) = \begin{cases}
        1, & \gamma_k < \tfrac{\pi}{3} - \tfrac{\pi}{30},\\
        0, & \gamma_k > \tfrac{\pi}{3} + \tfrac{\pi}{30},\\
        (-\gamma_k + (\tfrac{\pi}{3} + \tfrac{\pi}{30})) / (\tfrac{\pi}{15}),& \mbox{otherwise},
    \end{cases}
\end{equation*}
as shown in Figure \ref{fig:ex2_mu_fuzzy}.
Hence, for all $k \ge 0,$ it follows from \eqref{eq:wik},\eqref{eq:urk} that the control input calculated by the F-ARMA controller is given by
\begin{equation}
    u_{\rmr, k} = \frac{\mu_1 (\gamma_k) \ \sigma(\phi_{1,k} \theta_1) + \mu_2 (\gamma_k) \ \sigma(\phi_{2,k} \theta_2)}{\mu_1 (\gamma_k) + \mu_2 (\gamma_k)}.
\end{equation}
The ARMA controller coefficient vectors $\theta_1$ and $\theta_2$ are obtained by solving \eqref{eq:QP_1}, \eqref{eq:QP_2} with
$\ell_{\rmw,1} = 30,$ $\ell_{\rmw,2} = 10,$
\small
\begin{align*}
\SZ_1(r_k, y_k) &= \matl -p_k \\ \mbox{wrap}_\pi (\pi - \phi_k) \matr = \matl - \matl 1 & 0 \matr y_k \\ \mbox{wrap}_\pi (\pi - \matl 0 & 1 \matr y_k ) \matr,\\
\SZ_2(r_k, y_k) &=  \matl -p_k \\ \mbox{wrap}_\pi (- \phi_k) \matr = \matl - \matl 1 & 0 \matr y_k \\ \mbox{wrap}_\pi (- \matl 0 & 1 \matr y_k ) \matr,
\end{align*}
\normalsize
for all $k \ge 0,$
and $R_{\theta,1} = 10^{-3} I_{\ell_{\theta,1}},$ $R_{\theta,2} = 10^{-8} I_{\ell_{\theta,2}}.$
For $\theta_1,$ the $y$ and $u$ data obtained from the NMPC swing-up maneuver simulation results shown in Figure \ref{fig:ex2_train} corresponding to $t \in [0, 2.5]$ s is used for training, such that $\ell_{{\rm tr}, 1} = 126$ and, for all $k \in \{0, \ldots, \ell_{{\rm tr}, 1}-1\},$ $u_{{\rm tr}, 1, k} = u_{{\rm su}, k}$ and $y_{{\rm tr}, 1, k} = y_{{\rm su}, k}.$
For $\theta_2,$ the $y$ and $u$ data obtained from the NMPC stabilization near equilibrium simulation results shown in Figure \ref{fig:ex2_train} corresponding to $t \in [0, 15]$ s is used for training, such that $\ell_{{\rm tr}, 2} = 751$ and, for all $k \in \{0, \ldots, \ell_{{\rm tr}, 2}-1\},$ $u_{{\rm tr}, 2, k} = u_{{\rm st}, k}$ and $y_{{\rm tr}, 2, k} = y_{{\rm st}, k}.$

\begin{figure}[h!]
    \vspace{1em}
    \centering
    \includegraphics[width=0.9\columnwidth]{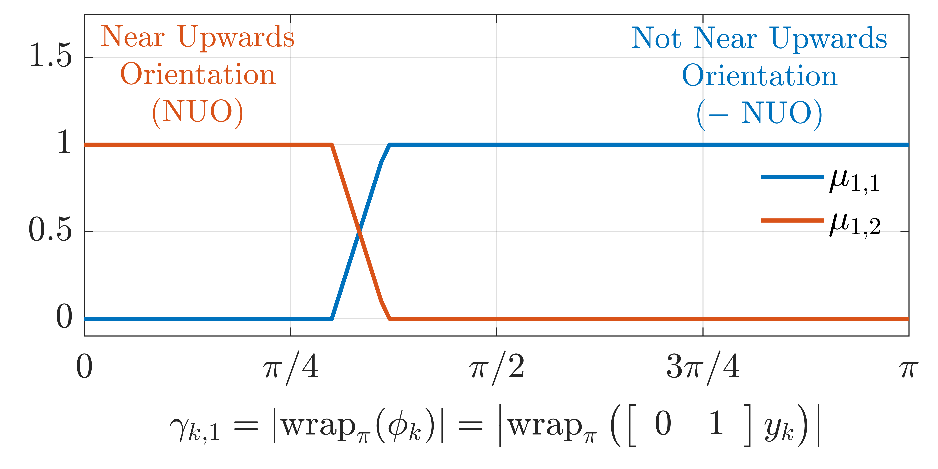}
    \vspace{-0.5em}
    \caption{Example \ref{ex:cart_pendulum}: {\bf Inverted Pendulum on Cart}. Fuzzy membership functions $\mu_1$ and $\mu_2$ corresponding to the fuzzy sets in which $\gamma_k$ is not near the upwards orientation ($-$ NUO) and near the upwards orientation (NUO), respectively.} 
    \label{fig:ex2_mu_fuzzy}
     \vspace{-1em}
\end{figure}

The results from implementing the trained F-ARMA controller on the inverted pendulum on a cart dynamics for $x(0) = \matl 0 & 0 & \tfrac{19}{20} \pi & 0\matr^\rmT$ are shown in Figure \ref{fig:ex2_yu}.
While the response of the F-ARMA controller shown in Figure \ref{fig:ex2_yu} is different from the response of the NMPC controller shown in Figure \ref{fig:ex2_train}, the F-ARMA controller is able to complete the swing-up maneuver using data obtained from the NMPC controller.
Furthermore, the average times taken to run NMPC and F-ARMA for the swing-up maneuver at each step are $1.1$ s and $5.4 \times 10^{-5}$ s, respectively.
\hfill {\LARGE$\diamond$}

\begin{figure}[h!]
    \centering
    \includegraphics[width=0.85\columnwidth]{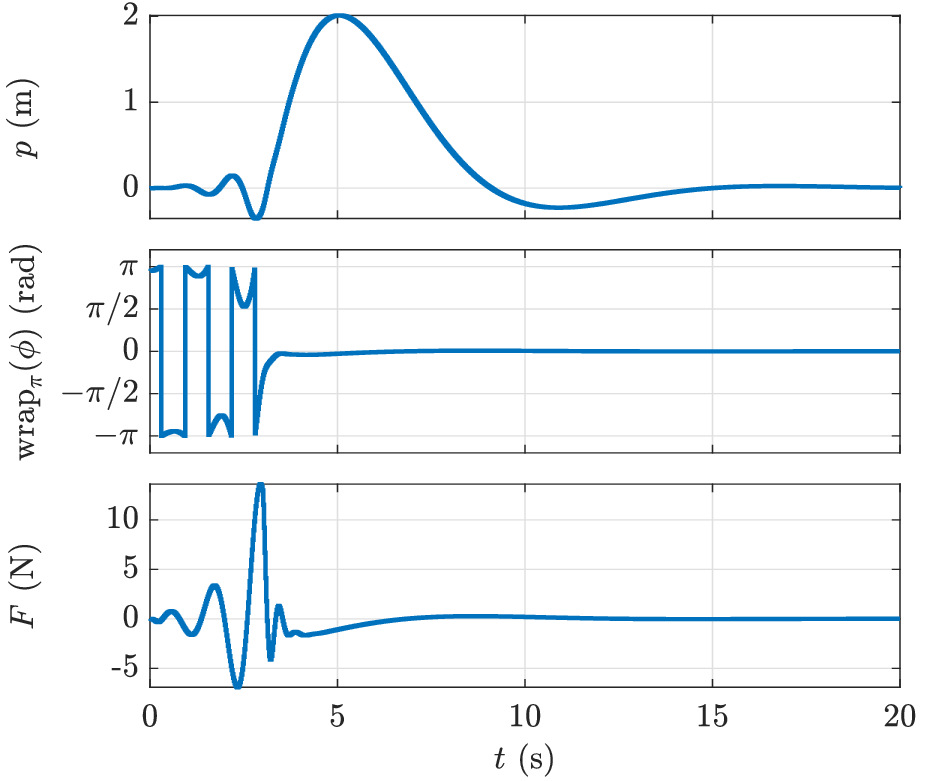}
    \vspace{-0.5em}
    \caption{Example \ref{ex:cart_pendulum}: {\bf Inverted Pendulum on Cart}. Results from implementing the trained F-ARMA controller on the inverted pendulum on cart dynamics given by \eqref{eq:xeq_ex2},\eqref{eq:yeq_ex2} for $x(0) = \matl 0 & 0 & \tfrac{19}{20} \pi & 0\matr^\rmT.$} 
    \label{fig:ex2_yu}
     \vspace{-2em}
\end{figure}
\end{exam}


\section{Conclusions} \label{sec:conclusions}
This paper presented an MPC-guided fuzzy controller synthesis framework called the F-ARMA controller. In this framework, ARMA controllers are trained using data obtained from MPC closed-loop simulations, and each ARMA controller emulates the response of the MPC controller under some desirable conditions.
The responses of all the trained ARMA controllers is then weighted depending on the measured system conditions and interpolated using a T-S fuzzy system.
%
Numerical examples illustrate the MPC-guided controller synthesis framework and show that the trained F-ARMA controller is able to emulate the constrained response of the original MPC controller.

%
Future work will focus on implementing and validating the F-ARMA framework in physical experiments and automating the choice of the fuzzy system rules and membership function parameters by leveraging fuzzy system optimization techniques\cite{wu2020,cui2022}.

\bibliographystyle{IEEEtran}
\bibliography{IEEEabrv,bib_paper}

\begin{thebibliography}{10}
\providecommand{\url}[1]{#1}
\csname url@samestyle\endcsname
\providecommand{\newblock}{\relax}
\providecommand{\bibinfo}[2]{#2}
\providecommand{\BIBentrySTDinterwordspacing}{\spaceskip=0pt\relax}
\providecommand{\BIBentryALTinterwordstretchfactor}{4}
\providecommand{\BIBentryALTinterwordspacing}{\spaceskip=\fontdimen2\font plus
\BIBentryALTinterwordstretchfactor\fontdimen3\font minus \fontdimen4\font\relax}
\providecommand{\BIBforeignlanguage}[2]{{%
\expandafter\ifx\csname l@#1\endcsname\relax
\typeout{** WARNING: IEEEtran.bst: No hyphenation pattern has been}%
\typeout{** loaded for the language `#1'. Using the pattern for}%
\typeout{** the default language instead.}%
\else
\language=\csname l@#1\endcsname
\fi
#2}}
\providecommand{\BIBdecl}{\relax}
\BIBdecl

\bibitem{grune2017}
L.~Gr{\"u}ne, J.~Pannek, L.~Gr{\"u}ne, and J.~Pannek, \emph{{Nonlinear Model Predictive Control}}.\hskip 1em plus 0.5em minus 0.4em\relax Springer, 2017.

\bibitem{rakovic2018}
S.~V. Rakovic and W.~S. Levine, \emph{Handbook of model predictive control}.\hskip 1em plus 0.5em minus 0.4em\relax Springer, 2018.

\bibitem{schwenzer2021}
M.~Schwenzer, M.~Ay, T.~Bergs, and D.~Abel, ``Review on model predictive control: {An} engineering perspective,'' \emph{Int. J. Adv. Manufact. Tech.}, vol. 117, no.~5, pp. 1327--1349, 2021.

\bibitem{mesbah2016}
A.~Mesbah, ``{Stochastic model predictive control: An overview and perspectives for future research},'' \emph{IEEE Cont. Syst. Mag.}, vol.~36, no.~6, pp. 30--44, 2016.

\bibitem{saltik2018}
M.~B. Salt{\i}k, L.~{\"O}zkan, J.~H. Ludlage, S.~Weiland, and P.~M. Van~den Hof, ``{An outlook on robust model predictive control algorithms: Reflections on performance and computational aspects},'' \emph{J. Proc. Contr.}, vol.~61, pp. 77--102, 2018.

\bibitem{alessio2009}
A.~Alessio and A.~Bemporad, ``A survey on explicit model predictive control,'' in \emph{Nonlinear Model Predictive Control: Towards New Challenging Applications}.\hskip 1em plus 0.5em minus 0.4em\relax Springer, 2009, pp. 345--369.

\bibitem{wen2009}
C.~Wen, X.~Ma, and B.~E. Ydstie, ``Analytical expression of explicit {MPC} solution via lattice piecewise-affine function,'' \emph{Automatica}, vol.~45, no.~4, pp. 910--917, 2009.

\bibitem{scibilia2009}
F.~Scibilia, S.~Olaru, and M.~Hovd, ``Approximate explicit linear {MPC} via {Delaunay} tessellation,'' in \emph{Proc. Europ. Contr. Conf.}\hskip 1em plus 0.5em minus 0.4em\relax IEEE, 2009, pp. 2833--2838.

\bibitem{jones2010}
C.~N. Jones and M.~Morari, ``Polytopic approximation of explicit model predictive controllers,'' \emph{IEEE Trans. Automat. Contr.}, vol.~55, no.~11, pp. 2542--2553, 2010.

\bibitem{kvasnica2011}
M.~Kvasnica and M.~Fikar, ``Clipping-based complexity reduction in explicit {MPC},'' \emph{IEEE Trans. Automat. Contr.}, vol.~57, no.~7, pp. 1878--1883, 2011.

\bibitem{nguyen2017}
N.~A. Nguyen, M.~Gulan, S.~Olaru, and P.~Rodriguez-Ayerbe, ``{Convex lifting: Theory and control applications},'' \emph{IEEE Trans. Automat. Contr.}, vol.~63, no.~5, pp. 1243--1258, 2017.

\bibitem{kvasnica2019}
M.~Kvasnica, P.~Bakar{\'a}{\v{c}}, and M.~Klau{\v{c}}o, ``{Complexity reduction in explicit MPC: A reachability approach},'' \emph{Syst. Contr. Lett.}, vol. 124, pp. 19--26, 2019.

\bibitem{zhang2016}
T.~Zhang, G.~Kahn, S.~Levine, and P.~Abbeel, ``Learning deep control policies for autonomous aerial vehicles with {MPC}-guided policy search,'' in \emph{Proc. Int. Conf. Robot. Automat.}\hskip 1em plus 0.5em minus 0.4em\relax IEEE, 2016, pp. 528--535.

\bibitem{hertneck2018}
M.~Hertneck, J.~K{\"o}hler, S.~Trimpe, and F.~Allg{\"o}wer, ``Learning an approximate model predictive controller with guarantees,'' \emph{IEEE Contr. Syst. Lett.}, vol.~2, no.~3, pp. 543--548, 2018.

\bibitem{kaufmann2020}
E.~Kaufmann, A.~Loquercio, R.~Ranftl, M.~M{\"u}ller, V.~Koltun, and D.~Scaramuzza, ``Deep drone acrobatics,'' \emph{arXiv preprint arXiv:2006.05768}, 2020.

\bibitem{zhang2020}
X.~Zhang, M.~Bujarbaruah, and F.~Borrelli, ``Near-optimal rapid {MPC} using neural networks: A primal-dual policy learning framework,'' \emph{IEEE Trans. Contr. Syst. Tech.}, vol.~29, no.~5, pp. 2102--2114, 2020.

\bibitem{chen2021}
H.~Chen, N.~Paoletti, S.~A. Smolka, and S.~Lin, ``{MPC}-guided imitation learning of bayesian neural network policies for the artificial pancreas,'' in \emph{Proc. Conf. Dec. Contr.}\hskip 1em plus 0.5em minus 0.4em\relax IEEE, 2021, pp. 2525--2532.

\bibitem{tagliabue2022}
A.~Tagliabue and J.~P. How, ``Output feedback tube {MPC-guided} data augmentation for robust, efficient sensorimotor policy learning,'' in \emph{Proc. Int. Conf. Intell. Rob. Syst.}\hskip 1em plus 0.5em minus 0.4em\relax IEEE, 2022, pp. 8644--8651.

\bibitem{tagliabue2024}
------, ``Tube-{NeRF}: {Efficient} imitation learning of visuomotor policies from {MPC} via tube-guided data augmentation and {NeRFs},'' \emph{IEEE Robot. Automat. Lett.}, 2024.

\bibitem{lilly2011}
J.~H. Lilly, \emph{Fuzzy control and identification}.\hskip 1em plus 0.5em minus 0.4em\relax John Wiley \& Sons, 2011.

\bibitem{nguyen2019}
A.-T. Nguyen, T.~Taniguchi, L.~Eciolaza, V.~Campos, R.~Palhares, and M.~Sugeno, ``Fuzzy control systems: {Past, present and future},'' \emph{IEEE Comp. Intell. Mag.}, vol.~14, no.~1, pp. 56--68, 2019.

\bibitem{precup2024}
R.-E. Precup, A.-T. Nguyen, and S.~Bla{\v{z}}i{\v{c}}, ``A survey on fuzzy control for mechatronics applications,'' \emph{Int. J. Syst. Sci.}, vol.~55, no.~4, pp. 771--813, 2024.

\bibitem{leal2021}
I.~S. Leal, C.~Abeykoon, and Y.~S. Perera, ``Design, simulation, analysis and optimization of {PID} and fuzzy based control systems for a quadcopter,'' \emph{Electronics}, vol.~10, no.~18, p. 2218, 2021.

\bibitem{wang2022}
Z.~Wang, K.~Sun, S.~Ma, L.~Sun, W.~Gao, and Z.~Dong, ``Improved linear quadratic regulator lateral path tracking approach based on a real-time updated algorithm with fuzzy control and cosine similarity for autonomous vehicles,'' \emph{Electronics}, vol.~11, no.~22, p. 3703, 2022.

\bibitem{al2023}
B.~M. Al-Hadithi, J.~M. Ad{\'a}nez, and A.~Jim{\'e}nez, ``A multi-strategy fuzzy control method based on the {Takagi-Sugeno} model,'' \emph{Opt. Contr. Appl. Meth.}, vol.~44, no.~1, pp. 91--109, 2023.

\bibitem{aslam2023}
S.~Aslam, Y.-C. Chak, M.~H. Jaffery, R.~Varatharajoo, and E.~A. Ansari, ``Model predictive control for {Takagi--Sugeno} fuzzy model-based spacecraft combined energy and attitude control system,'' \emph{Adv. Space Res.}, vol.~71, no.~10, pp. 4155--4172, 2023.

\bibitem{mendes2024}
T.~P.~G. Mendes, A.~M. Ribeiro, L.~Schnitman, and I.~B. Nogueira, ``A {PLC}-embedded implementation of a modified {Takagi--Sugeno--Kang}-based {MPC} to control a pressure swing adsorption process,'' \emph{Processes}, vol.~12, no.~8, p. 1738, 2024.

\bibitem{sayadian2024}
N.~Sayadian, F.~Jahangiri, and M.~Abedi, ``Adaptive event-triggered fuzzy {MPC} for unknown networked {IT-2} {TS} fuzzy systems,'' \emph{Int. J. Dyn. Contr.}, pp. 1--20, 2024.

\bibitem{cervantes2020}
J.~S. Cervantes-Rojas, F.~Mu{\~n}oz, I.~Chairez, I.~Gonz{\'a}lez-Hern{\'a}ndez, and S.~Salazar, ``Adaptive tracking control of an unmanned aerial system based on a dynamic neural-fuzzy disturbance estimator,'' \emph{ISA Trans.}, vol. 101, pp. 309--326, 2020.

\bibitem{pham2023}
D.-H. Pham, C.-M. Lin, V.-P. Vu, H.-Y. Cho \emph{et~al.}, ``Design of missile guidance law using {Takagi-Sugeno-Kang} (tsk) elliptic type-2 fuzzy brain imitated neural networks,'' \emph{IEEE Access}, vol.~11, pp. 53\,687--53\,702, 2023.

\bibitem{khater2024}
A.~A. Khater, E.~M. Gaballah, M.~El-Bardin, and A.~M. El-Nagar, ``Real time adaptive probabilistic recurrent {Takagi-Sugeno-Kang} fuzzy neural network proportional-integral-derivative controller for nonlinear systems,'' \emph{ISA Trans.}, vol. 152, pp. 191--207, 2024.

\bibitem{boggs1995}
P.~T. Boggs and J.~W. Tolle, ``Sequential quadratic programming,'' \emph{Acta Numerica}, vol.~4, pp. 1--51, 1995.

\bibitem{fernandez2010}
D.~Fern{\'a}ndez and M.~Solodov, ``Stabilized sequential quadratic programming for optimization and a stabilized {Newton-type} method for variational problems,'' \emph{Math. Prog.}, vol. 125, pp. 47--73, 2010.

\bibitem{izmailov2012}
A.~F. Izmailov and M.~V. Solodov, ``Stabilized {SQP} revisited,'' \emph{Math. Prog.}, vol. 133, no.~1, pp. 93--120, 2012.

\bibitem{andersson2019}
J.~A.~E. Andersson, J.~Gillis, G.~Horn, J.~B. Rawlings, and M.~Diehl, ``{CasADi} -- {A} software framework for nonlinear optimization and optimal control,'' \emph{Mathematical Programming Computation}, vol.~11, no.~1, pp. 1--36, 2019.

\bibitem{cart_pendulum_MPC}
J.~A. Paredes~Salazar, ``{MPC for cart pendulum stabilization},'' \url{https://github.com/JAParedes/MPC-for-cart-pendulum-stabilization}, accessed: 09-29-2024.

\bibitem{wu2020}
D.~Wu and X.~Tan, ``Multitasking genetic algorithm {(MTGA)} for fuzzy system optimization,'' \emph{IEEE Trans. Fuzzy Syst.}, vol.~28, no.~6, pp. 1050--1061, 2020.

\bibitem{cui2022}
Y.~Cui, Y.~Xu, R.~Peng, and D.~Wu, ``Layer normalization for {TSK} fuzzy system optimization in regression problems,'' \emph{IEEE Trans. Fuzzy Syst.}, vol.~31, no.~1, pp. 254--264, 2022.

\end{thebibliography}

\end{document}